\documentclass[12pt]{article}

\catcode`\@=11
\global\arraycolsep=2pt
\usepackage{amsbsy,amssymb,latexsym,amsfonts,amsmath}
\usepackage{graphicx,color}

\allowdisplaybreaks

\begin{document}
\begin{flushright}
\parbox{4.2cm}
{RUP-22-19}
\end{flushright}

\vspace*{0.7cm}

\begin{center}
{ \Large Geometrical Trinity of Unimodular Gravity}
\vspace*{1.5cm}\\
{Yu Nakayama}
\end{center}
\vspace*{1.0cm}
\begin{center}

Department of Physics, Rikkyo University, Toshima, Tokyo 171-8501, Japan

\vspace{3.8cm}
\end{center}

\begin{abstract}
We construct a Weyl transverse diffeomorphism invariant theory of teleparallel gravity by employing the Weyl compensator formalism. The low-energy dynamics has a single spin two gravition without a scalar degree of freedom. By construction, it is equivalent to unimodular gravity (as well as Einstein's general relativity with an adjustable cosmological constant) at the non-linear level. Combined with our earlier construction of a Weyl transverse diffeomorphism invariant theory of symmetric teleparallel gravity,  unimodular gravity is represented in three alternative ways.
\end{abstract}

\thispagestyle{empty} 

\setcounter{page}{0}

\newpage

\section{Introduction}
The realization of massless particles with higher spin in field theories necessary involves redundancy in their description because the number of components in fields is much larger than the physical degrees of freedom \cite{Weinberg:1980kq}. In order to remove the redundancy, we typically assume a gauge symmetry of the action and employ the gauge principle. The necessary gauge symmetry may not be unique. In the field theory with massless spin two particles, there are two minimal possibilities: the one is to use the diffeomorphism and the other is to use the Weyl transverse diffeomorphism \cite{Alvarez:2006uu}.\footnote{The Weyl transverse diffeomorphism is a combination of the Weyl transformation of the metric and the diffeomorphism whose Jacobian is unity.}

The former description of the massless spin-two particles is well-known. It gives Einstein's general relativity at the non-linear level \cite{Deser:1969wk}. The latter description of the massless spin-two particles seems less known, but it gives unimodular gravity at the non-linear level. It turns out that, at least classically, both descriptions are equivalent as we will review in the main text (see e.g. \cite{Alvarez:2005iy}\cite{Carballo-Rubio:2022ofy} for a review). 

Einstein's general relativity and unimodular gravity are based on the (pseudo) Riemannian geometry, where the diffeomorphism (as well as Weyl symmetry) is naturally defined. It is, however, possible to reformulate the gravitational theory in terms of other geometries. For example, it is possible to formulate the equivalent gravitational theory in teleparallel geometry (e.g. \cite{tele}\cite{Maluf:2013gaa}\cite{Bahamonde:2021gfp} and reference therein) or symmetric teleparallel geometry \cite{Nester:1998mp}\cite{BeltranJimenez:2017tkd}, where the connection is torsional or metric incompatible respectively. While the appearance is different, the dynamical contents turn out to be equivalent by choosing an appropriate action \cite{Jimenez:2019woj}\cite{BeltranJimenez:2019odq}\cite{Bahamonde:2021gfp}\cite{Lu:2021wif}.

In this paper, we construct a Weyl transverse diffeomorphism invariant theory of teleparallel gravity by employing the Weyl compensator formalism. We show that it is equivalent to unimodular gravity (as well as Einstein's general relativity with an adjustable cosmological constant) at the non-linear level. Combined with our earlier construction of a Weyl transverse diffeomorphism invariant theory of symmetric teleparallel gravity \cite{Nakayama:2021rda}, unimodular gravity is represented in three alternative ways. Our systematic construction offers a unified framework to discuss various representations of massless spin-two particles with different gauge symmetry. 

The organization of our paper is as follows. In section 2, we construct a Weyl transverse diffeomorphism invariant theory of teleparallel gravity and show it is equivalent to unimodular gravity. In section 3, we first review a Weyl transverse diffeomorphism invariant theory of symmetric teleparallel gravity and offer a unified framework to discuss these theories. In section 4, we conclude with some discussions.

\section{Weyl transverse diffeomorphism invariant theory of teleparallel gravity}
Teleparallel gravity is formulated in terms of a metric-compatible but torsional flat connection $\Gamma^{\mu}_{\ \rho\sigma}$ (where no symmetry of the lower indices is assumed). In order to understand the dynamical degrees of freedom, it is convenient to use the tetrad formalism by introducing the tetrad $e_{\mu}^a$ and an independent spin connection $\omega^{a}_{\ b \mu}$. The metric can be reconstructed from the tetrad as $g_{\mu\nu} = \eta_{ab}e^a_\mu e^b_\nu$, where $\eta_{ab} = \mathrm{diag}(-1,1,1,1)$. 

The two connections are related through the tetrad compatibility condition:
\begin{align}
\partial_\mu e^{a}_\nu + \omega^a_{\ b \mu} e^{b}_\nu -\Gamma^{\rho}_{\ \mu \nu} e^{a}_\rho = 0 \ . 
\end{align}
Thanks to the tetrad compatibility condition, the tensor indices are raised, lowered and converted by using the metric $g_{\mu\nu}$, the inverse metric $g^{\mu\nu}$, the tetrad $e^a_{\mu}$ and the inverse tetrad $e_a^\mu$ as usual in the general relativity.
The curvature tensor is given by 
\begin{align}
R^{a}_{\ b \mu\nu} = \partial_\mu \omega^{a}_{\ b \nu} - \partial_\nu \omega^a_{\ b \mu} + \omega^a_{\ c \mu} \omega^c_{\ b \nu} -  \omega^a_{\ c \nu} \omega^c_{\ b \mu} \ , \label{curvature}
\end{align}
and we assume that it vanishes in teleparallel geometry.

Instead of the curvature, in teleparallel gravity, we assume that the gravitational force is propagated by a torsion.
In terms of the spin connection and the tetrad, the torsion tensor is defined by
\begin{align}
T^a_{\ \mu\nu} = \partial_\mu e^a_\nu - \partial_\nu e^{a}_\mu +\omega^{a}_{\ b \mu} e^{b}_\nu - \omega^{a}_{\ b \nu} e^{b}_\mu  \ .  \label{torsion}
\end{align}
 We also define $T_\nu = e_a^\mu T^a_{\ \mu \nu}$ for later notational convenience. The assumption of the flat connection (i.e. vanishing of the curvature tensor) means that we can (locally) choose the  Weitzenb\"ock gauge, where $\omega^a_{\ b \mu} = 0$. In some literature on teleparallel gravity, this gauge condition for the local Lorentz transformation is implicitly assumed from the beginning.
 
 Let us introduce a particular quadratic combination of the torsion tensor
 \begin{align}
T_E= \frac{1}{4} T_{\mu\nu\rho} T^{\mu\nu\rho} +\frac{1}{2} T_{\mu\nu\rho} T^{\mu\rho\nu} - T_\mu T^\mu \ . 
 \end{align}
 With or without imposing the Weitzenb\"ock gauge, vanishing of the curvature  tensor leads to the identity for the Ricci scalar $\mathcal{R}$, which is constructed out of the metric through the Christoffel connection$\{^\mu_{\ \rho \sigma}\} = \frac{1}{2} g^{\mu \nu} (\partial_\rho g_{\nu\sigma} + \partial_\sigma g_{\nu \rho} - \partial_\nu g_{\rho \sigma}) $:
 \begin{align}
- \mathcal{R} &= - g^{\mu\nu} \left( \partial_\lambda \{^\lambda_{\ \mu \nu}\} - \partial_\nu \{^\lambda_{\ \mu \lambda}\} + \{^\sigma_{\ \mu \nu}\} \{^\lambda_{\ \lambda \sigma}\} - \{^\sigma_{\ \mu \lambda}\} \{^\lambda_{\ \nu \sigma}\} \right) \cr
              &=T_E -2 e^{-1} \partial_\mu (e T^\mu) \ , 
             \label{identity}
 \end{align}
 where $e = g^{\frac{1}{2}} =  \sqrt{|\mathrm{det}g_{\mu\nu}|}$.
 This identity is the starting point for the so-called teleparallel equivalent of general relativity \cite{tele}.

Our first mission is to construct a teleparallel equivalent of unimodular gravity, and for this purpose, it is convenient to start with the action that is invariant under Weyl transformation.\footnote{In the literature, it is also called ``conformal  transformation", but we preferably use the word Weyl transformation in this paper. It is also different from the scale transformation \cite{Nakayama:2013is}.} This problem was addressed in \cite{Maluf:2011kf}\cite{Bamba:2013jqa}\cite{Momeni:2014taa} and we closely follow their approach. 

Let us first define the Weyl transformation of the torsion tensor. By definition, the tetrad transforms as $\delta_W e_{\mu}^a = \sigma  e_{\mu}^a$ under the infinitesimal Weyl transformation (so that $\delta_W e = 4 \sigma e$ in four dimensions). The transformation of the connection under the Weyl transformation may have some arbitrariness (as in the Palatini formalism \cite{Edery:2019txq}) in more general situations, but here it must be compatible with the flatness condition, and it fixes the transformation law. In practice, we can use the decomposition of the torsion tensor \eqref{torsion} in the Weitzenbl\"ock gauge and study its Weyl variation:
\begin{align}
\delta_W T^{a}_{\ \mu\nu} &= \partial_\mu \delta_W e^a_\nu -     \partial_\nu \delta_W e^a_\mu  \cr
& = \sigma T^{a}_{\ \mu\nu} + (\partial_\mu \sigma) e_\nu^a - (\partial_\nu \sigma) e_\mu^a \ .  
\end{align}
Equivalently, we have assumed $\delta_W \omega^{a}_{\ b \mu} = 0$ in more general gauge. Note that under this assumption, the curvature tensor \eqref{curvature} is invariant under the Weyl transformation (unlike the Riemann curvature tensor), and the flatness condition is compatible with the Weyl symmetry.

Accordingly, the quadratic scalars constructed out of the torsion tensor transform as
\begin{align}
\delta_W (T^{\mu\nu\rho}T_{\mu\nu\rho}) &= -2\sigma (T^{\mu\nu\rho}T_{\mu\nu\rho}) - 4 T^\mu \partial_\mu \sigma \cr
\delta_W (T^{\mu\nu\rho}T_{\nu\mu\rho}) &= -2\sigma (T^{\mu\nu\rho}T_{\nu\mu\rho}) - 2 T^\mu \partial_\mu \sigma \cr
\delta_W (T^{\mu}T_{\mu}) &= -2\sigma (T^{\mu}T_{\mu}) - 6 T^\mu \partial_\mu \sigma  \ 
\end{align}
under the infinitesimal Weyl transformation (in four dimensions). We also recall that the Ricci scalar transforms under the Weyl transformation as
\begin{align}
\delta_W \mathcal{R} = -2\sigma \mathcal{R} -6 \Box \sigma \  
\end{align}
in four dimensions. Here the connection used in the Laplacian $\Box$ is the Christoffel connection.

To construct a Weyl invariant action, we introduce the Weyl compensator $\phi$, which transforms under the infinitesimal Weyl transformation 
as
\begin{align}
\delta_W \phi = -\sigma \phi \ . 
\end{align}
The compensator is a scalar under diffeomorphism. It will be non-dynamical, and we use it to fix a part of the gauge symmetry.

Let us now construct the Weyl invariant action. 
We regard the Weyl symmetry as a gauge symmetry in the sense that the solutions related by the Weyl transformation are physically equivalent. We also demand the diffeomorphism and the Local Lorentz symmetry as gauge symmetries. Up to two derivatives, the candidate action is
\begin{align}
S_W = \int d^4x &e  \phi^2\left( A T^{\mu\nu\rho} T_{\mu\nu\rho} + B T^{\mu\nu\rho} T_{\nu \mu \rho} + C T^\mu T_{\mu}  \right)  \cr 
&+ e \xi g^{\mu\nu} (\partial_\mu -\frac{1}{3}T_{\mu})\phi (\partial_\nu -\frac{1}{3}T_{\nu})\phi + e\lambda \phi^4 \ .  \label{first}
\end{align}
The Weyl invariance demands $2A+ B +3C=0$, but otherwise we seem to have three arbitrary parameters (up to overall factors that can be absorbed by a definition of $\phi$).

In order to construct a consistent field theory of a massless spin-two  particle, however, we need to choose a particular parameter set $A = -\frac{1}{4}$, $B=-\frac{1}{2} $, $C=\frac{1}{3}$, $\xi = 6$ (up to overall rescaling). This choice was made in \cite{Maluf:2011kf}, and it leads to the teleparallel equivalent of general relativity by fixing the Weyl gauge symmetry with the condition $\phi=1$. Indeed, if we substitute $\phi=1$ into \eqref{first}, the action becomes 
\begin{align}
S_E &= \int d^4x e (-T_E + \lambda) \cr
& \int d^4x e(\mathcal{R} + \lambda) \ 
\end{align}
due to the identity \eqref{identity} (up to surface terms). If we take $e_\mu^a$ as a dynamical variable, we reproduce the Einstein-Hilbert action with a cosmological constant.

Why do we have to take this particular parameter set among other Weyl invariant sets? 
There are several different viewpoints about this particular choice of action. 

If we use the Weitzenb\"ock gauge $\omega^a_{\ \mu\nu} = 0$, the local Lorentz symmetry was already fixed, and there is generically no reason to expect that the action written in terms of the tetrad (in the Weitzenb\"ock gauge) is invariant under the local Lorentz transformation any longer. Nevertheless, the particular choice above leads to the emergent local Lorentz symmetry (up to a surface term) even in the Weitzenb\"ock gauge. This emergent local Lorentz symmetry can be used to remove unphysical degrees of freedom and achieve the equivalence to Einstein's general relativity.

If we do not use the Weitzenb\"ock gauge, the relevance of this particular parameter choice can be seen as a decoupling of the spin connection from the action (again up to surface terms). Thus, the spin connection is not a dynamical degree of freedom in this particular theory, and the dynamical contents become identical to Einstein's general relativity (by discarding the spin connection that does not appear in the action).

From now on, we would like to construct a Weyl transverse diffeomorphism invariant theory of teleparallel gravity, which will be our first main goal of this paper. The transverse diffeomorphism is a subset of the diffeomorphism whose Jacobian is unity: $\det(\partial_\mu \xi^\nu) = 1$. When it is also invariant under the Weyl symmetry, such a gravitational theory describes a massless spin-two particle at the linearized level. We will show the consistency at the non-linear level by showing the equivalence to unimodular gravity (and Einstein's general relativity with an adjustable cosmological constant).

The starting point is the action given by \eqref{first}, which is invariant under full diffeomorphism and Weyl transformation. Instead of fixing the Weyl symmetry, we are going to fix the volume-changing part of the diffeomorphism by setting $\phi = e^{-\frac{1}{4}}$. The resulting action becomes
\begin{align}
S_U = \int d^4x &e^{\frac{1}{2}}\left( -\frac{1}{4} T^{\mu\nu\rho} T_{\mu\nu\rho} -\frac{1}{2} T^{\mu\nu\rho} T_{\nu \mu \rho} + \frac{1}{3} T^\mu T_{\mu}  \right)  \cr
& + e 6 g^{\mu\nu} (\partial_\mu -\frac{1}{3}T_{\mu})e^{-\frac{1}{4}} (\partial_\nu -\frac{1}{3}T_{\nu})e^{-\frac{1}{4}} + \lambda \label{second} \ .
\end{align}
Note that $\lambda$ term is just constant (without depending on $e^a_\mu$ unlike the cosmological constant in the Einstein-Hilbert action) and we may simply discard it as a classical action.

By using identity \eqref{identity} and integration by part, we can further rewrite the action into the form:
\begin{align}
S_U = \int d^4x e^{\frac{1}{2}}\left(\mathcal{R} + \frac{3}{32}g^{\mu\nu} \partial_\mu g \partial_\nu g \right) \ ,
\end{align}
where we emphasize again that $\mathcal{R}$ is the standard Ricci scalar constructed out of metric tensor through the Christoffel connection.
This action is invariant under the Weyl transformation and the transverse diffeomorphism (i.e. diffeomorphism whose Jacobian is unity). One may further fix the Weyl symmetry by setting e.g. $e=g^{\frac{1}{2}}=1$, but we opt not to (because the variation of $g_{\mu\nu}$ becomes non-trivial after imposing the non-linear constraint such as $e=g^{\frac{1}{2}}=1$). It describes unimodular gravity by regarding $e_a^\mu$ as dynamical variable \cite{Alvarez:2006uu}\cite{Alvarez:2010cg}\cite{Oda:2016pok}.
 Since the cosmological constant $\lambda$ does not appear in the teleparallel equivalent of unimodular gravity, it is equivalent to Einstein's general relativity with an adjustable cosmological constant \cite{Henneaux:1989zc}. Indeed, it effectively appears as a boundary condition in the equations of motion.

Alternatively, one may linearize \eqref{second} around the Minkowski space $e^a_\mu = \delta^a_\mu + \eta^{a\nu} \frac{1}{2} h_{\nu \mu}$ (so that $g_{\mu\nu} = \eta_{\mu\nu} + h_{\mu\nu}$). Due to the (emergent) local Lorentz symmetry, only the symmetric part of $h_{\mu\nu}$ contributes to the action.
Explicitly, the linearized action is given by 
\begin{align}
S_U = \int d^4x \left(- \frac{1}{4} \partial_\alpha h_{\mu\nu} \partial^\alpha h^{\mu\nu} +\frac{1}{2} \partial_\mu h^{\mu\alpha} \partial^\nu h_{\nu\alpha} - \frac{1}{4}\partial_\mu h \partial_\nu h^{\nu\mu} +\frac{3}{32}\partial_\mu h \partial^\mu h  \right) \ , 
\end{align}
where $h_{\mu\nu} = h_{\nu\mu}$ is assumed, and $h = \eta^{\mu\nu} h_{\mu\nu}$. The indices are raised and lowered by the Minkowski metric $\eta_{\mu\nu}$ (only) here. We see that the linearized action can be expressed only in terms of  the traceless mode $\hat{h}_{\mu\nu} = h_{\mu\nu} - \frac{\eta_{\mu\nu}}{4} h $ and the linearized Weyl symmetry, a shift of the trace part of $h_{\mu\nu}$ becomes obvious. It is invariant under the transverse diffeomorphism $\delta h_{\mu\nu} = \partial_\mu \xi_\nu + \partial_\nu \xi_\mu$ with $\partial_\mu \xi^\mu = 0$.
As is known in the literature (e.g. see \cite{Alvarez:2006uu}), it describes a massless spin-two graviton without a scalar degree of freedom.

\section{Geometrical trinity of unimodular gravity}
In the previous section, we have constructed the teleparallel equivalent of the unimodular gravity. In our earlier work \cite{Nakayama:2021rda}, we have constructed the symmetric teleparallel equivalent of unimodular gravity. The situation is similar to the geometrical trinity of the general relativity presented in \cite{Jimenez:2019woj}. Our two constructions are both equivalent to unimodular gravity, and we may declare that we have completed the geometrical trinity of unimodular gravity. 

To make our presentation self-contained, let us briefly summarize the symmetric teleparallel equivalent of unimodular gravity and then offer a unified framework. In the symmetric teleparallel gravity, we assume  that the connection is flat and torsion-free but is not compatible with the metric. The dynamical degrees of freedom are encoded in the non-metricity tensor $Q^\mu_{\ \rho \sigma} = \nabla^\mu g_{\rho\sigma}$. The analogue of the crucial identity \eqref{identity} is   \begin{align}
-\mathcal{R} &=  Q_E  + g^{-\frac{1}{2}} \partial_ \mu(g^{\frac{1}{2}} Q^\mu -g^{\frac{1}{2}}\bar{Q}^\mu)  \cr
Q_E &= \frac{1}{4} Q_{\alpha \mu \nu} Q^{\alpha \mu \nu} -\frac{1}{{2}} Q_{\alpha \mu \nu} Q^{\mu \alpha \nu} -\frac{1}{4} Q_{\alpha} Q^{\alpha} + \frac{1}{2} \bar{Q}_\alpha Q^{\alpha} \ ,
\end{align}
where $\mathcal{R}$ is the Ricci scalar constructed out of the metric tensor through the Christoffel connection. 
Here we have introduced $Q_\mu = Q_{\mu \  \alpha}^{\ \alpha} $ and $\bar{Q}_{\mu} =  Q_{\ \alpha\mu }^{\alpha }$. 

Since the symmetric teleparallel connection is flat, we can always (at least locally) choose the so-called coincident gauge \cite{BeltranJimenez:2017tkd}\cite{BeltranJimenez:2022azb}, where $Q_{\alpha \mu\nu} = \partial_\alpha g_{\mu\nu}$.
Sticking with this gauge, one can define the infinitesimal Weyl transformation\cite{Nakayama:2021rda} (see  \cite{Iosifidis:2018zwo}\cite{Gakis:2019rdd} as well):
\begin{align}
\delta_W g_{\mu\nu} &= 2\sigma g_{\mu\nu} \cr
\delta_W g & = 8\sigma g \cr
\delta_W Q^{\alpha}_{\ \mu\nu} &= 2 g_{\mu\nu} g^{\alpha \beta} \partial_\beta \sigma \cr
\delta_W Q_\mu &= 8 \partial_\mu \sigma \cr
\delta_W \bar{Q}_\mu & = 2 \partial_\mu \sigma \cr
\delta_W \phi &= - \sigma \phi \ .  \label{Weyl}
\end{align}

With these transformations in mind, we may start with the Weyl invariant action
\begin{align}
S_W = \int d^4x \sqrt{g} \left( -\phi^2 (Q_E +  g^{-\frac{1}{2}} \partial_\mu(g^{\frac{1}{2}} Q^\mu -g^{\frac{1}{2}}\bar{Q}^\mu)  + 6 g^{\mu\nu} \partial_\mu \phi \partial_\nu \phi + \lambda  \phi^4 \right)  \ .  \label{start}
\end{align}
If we fix the Weyl symmetry by setting $\phi=1$, we obtain the Einstein-Hilbert action with a cosmological constant. Instead,
 we now fix the volume-changing part of the diffeomorphism by setting $\phi = g^{-\frac{1}{8}}$. The resulting Weyl transverse diffeomorphism invariant action is
\begin{align}
S_U = \int d^4x {g}^{\frac{1}{4}} (-{Q}_W) \ , 
\end{align}
where 
\begin{align}
{Q}_W = c_1 Q_{\alpha \mu \nu} Q^{\alpha \mu \nu} + c_2 Q_{\alpha \mu \nu} Q^{\mu \alpha \nu} + c_3 Q_{\alpha} Q^{\alpha} + c_4 \bar{Q}_{\alpha} \bar{Q}^{\alpha} + c_5 \bar{Q}_\alpha Q^{\alpha} \label{final}
\end{align}
with $c_1 = \frac{1}{4}$, $c_2= -\frac{1}{2}$, $c_3 = -\frac{3}{32}$, $c_4=0 $ and $c_5= \frac{1}{4}$. As in the previous section, the $\lambda$ term is dropped because it does not affect the equations of motion. This action describes symmetric teleparallel equivalent of unimodular gravity. By construction it is equivalent to unimodular gravity or Einstein's general relativity with an adjustable cosmological constant.

The teleparallel equivalent of unimodular gravity and the symmetric teleparallel equivalent of unimodular gravity can be unified in the following way.
We now consider the most generic connection which is flat but with torsion and non-metricity. Under the flatness condition, we have the identity
\begin{align}
-\mathcal{R} &= G_E + g^{-\frac{1}{2}}\partial_\mu \left(g^{\frac{1}{2}}(Q^\mu -\bar{Q}^\mu -2 T^\mu)\right) \cr
G_E& = T_E + Q_E + Q_{\mu\nu\rho} T^{\rho \mu\nu}-Q_\mu T^\mu + \bar{Q}_\mu T^\mu \ .
\label{ient3}
\end{align}
Starting with a particular form of the Weyl invariant action
\begin{align}
S_W = \int d^4x \sqrt{g} \left(-\phi^2 (G_E +  g^{-\frac{1}{2}}\partial_\mu \left(g^{\frac{1}{2}}(Q^\mu -\bar{Q}^\mu -2 T^\mu)\right) + 6 g^{\mu\nu}\partial_\mu \phi \partial_\nu \phi + \lambda \phi^4\right) \ , \label{Weyl2}
\end{align}
and fixing the gauge by setting $\phi = g^{-\frac{1}{8}}$, we obtain
\begin{align}
S_U = \int d^4x e^{\frac{1}{2}}\left(\mathcal{R} + \frac{3}{32}g^{\mu\nu} \partial_\mu g \partial_\nu g \right)  \ .  \label{generaluni}
\end{align}

Note that if we write the action in terms of $g_{\mu\nu}$ (or $e^a_\mu$) by using \eqref{ient3}, it is obvious that the resultant theory is unimodular gravity with the Weyl transverse diffeomorphism, but it is worth while mentioning what happens to the other degrees of freedom in the most generic  (flat) connection. For this purpose, we can use the flatness condition to parameterize the torsion and the non-metricity as 
\begin{align}
T^{\alpha}_{\ \mu \nu} &= (\Lambda^{-1})^\alpha_{\ \rho}(\partial_\mu \Lambda^\rho_{\ \nu} - \partial_\nu \Lambda^{\rho}_{\ \mu})  \cr
Q_{\alpha \mu\nu} &= \partial_\alpha g_{\mu\nu} - (\Lambda^{-1})^\lambda_{\ \rho}(\partial_\alpha \Lambda^{\rho}_{\ \mu} g_{\nu\lambda} + \partial_\alpha \Lambda^{\rho}_{\ \nu} g_{\mu\lambda} ) \ .
\end{align}
Importantly, $\Lambda^\rho_{\ \mu}$ does not appear in the action  given by \eqref{generaluni} (up to surface terms), so one may claim that the degrees of freedom encoded in $\Lambda^\rho_{\ \mu}$ is a ``gauge" degree of freedom. This is the enhanced symmetry, which makes the action \eqref{generaluni} special compared with the other general quadratic action of the torsion and the non-metricity (under the additional Weyl symmetry).

One may use the enhanced symmetry to set either $T^\alpha_{\ \mu\nu}=0$ or $Q_{\alpha \mu\nu}=0$ and then we have the teleparallel equivalent of  unimodular gravity and the symmetric teleparallel equivalent of unimodular gravity, and the construction here gives a unified framework. Note that the resultant theory still enjoys the Weyl transverse diffeomorphism necessary for unimodular gravity.

Let us briefly discuss the matter coupling in the (symmetric) teleparallel equivalent of unimodular gravity. It might occur to us that  we can use the covariant derivative that are natural in each geometry (i.e. the connection with torsion or the connection with non-metricity). It turns out, however, that such a choice would not be equivalent to the conventional matter coupling in unimodular gravity; what is more worse, it would be inconsistent. The reason is that we have emphasized that it is important to retain emergent local Lorentz or diffeomorphism symmetry even after imposing the Weitzenb\"ock gauge in teleparallel case or the coincident gauge in symmetric teleparallel case. Among many other possibilities, we have chosen the gravity action as \eqref{first} and \eqref{start} based on this requirement, and so should be in the matter action. We do not have to treat each case separately because the above construction gives the unified framework based on \eqref{Weyl2}.

Let us take some examples motivated from the standard model of particle physics. We begin with the gauge field. The gauge kinetic term as well as theta terms are Weyl invariant in four-dimensions, so
\begin{align}
S_{\mathrm{gauge}} = \int d^4x e \left(-\frac{1}{4 g^2} \mathrm{Tr} F^{\mu\nu}F_{\mu\nu} + \theta  \epsilon_{\mu\nu\rho\sigma} \mathrm{Tr} F^{\mu\nu} F^{\rho\sigma} \right)
\end{align}
does not change after fixing the gauge by $\phi=e^{-\frac{1}{4}}$. Here the field strength is defined by $F_{\mu\nu} = \partial_\mu A _\nu - \partial_\nu A_\mu + i[A_\mu,A_\nu]$, and its Weyl weight is zero. In the Riemannian geometry, one may replace the partial derivative with the covariant derivative in the definition of the field strength, but  here we cannot. To state it more explicitly, the coupling to the connection, in particular torsional part, is not allowed.

Next, let us consider the scalar field. A canonical example is the Higgs field. We start with the Weyl invariant action
\begin{align}
S_{\mathrm{scalar}} = -\int d^4x e \left(g^{\mu\nu}\partial_\mu \Phi \partial_\nu \Phi +\frac{1}{6}\mathcal{R} \Phi^2 + \phi^4 V(\Phi/\phi) \right) \ ,
\end{align}
where $\delta_W \Phi = -\sigma \Phi$.
Setting $\phi = e^{-\frac{1}{4}}$ to fix the gauge, we obtain the matter action compatible with Weyl transverse diffeomorphism.
\begin{align}
S_{\mathrm{scalar}} = -\int d^4x e \left(g^{\mu\nu}\partial_\mu \Phi \partial_\nu \Phi +\frac{1}{6}\mathcal{R} \Phi^2 + e^{-1} V(\Phi e^{\frac{1}{4}}) \right) \ . 
\end{align}
Note that the peculiar appearance of $e$ in the potential is necessary for the consistency of the gravitational equations of motion in unimodular gravity.

One may wonder if the non-minimal coupling is fixed here, but it is not the case. The non-minimal coupling can be introduced if we add a more elaborate Weyl invariant action
\begin{align}
S_{\text{non-minimal}} = \int d^4x e \xi_\Phi \left(-\frac{\Phi^2}{\phi^2} \phi \Box \phi +\frac{1}{6}\mathcal{R} \Phi^2 \right) \ .
\end{align}
to the above. Clearly, if we set $\phi=1$, we obtain the non-minimal coupling $R \Phi^2$ in Einstein's general relativity. If we instead set $\phi=e^{-\frac{1}{4}}$, we obtain the equivalent non-minimal coupling in unimodular gravity:
\begin{align}
S_{\text{non-minimal}} = \int d^4x e \xi_\Phi \left(-\Phi^2 e^{\frac{1}{4}} \Box e^{-\frac{1}{4}} +  \frac{1}{6}\mathcal{R} \Phi^2 \right) \ .
\end{align}

Finally, consider the Weyl invariant fermion action (where $\delta_W \Psi = -\frac{3}{2} \sigma \Psi$)
\begin{align}
S_{\mathrm{fermion}} = \int d^4x e \left( i\bar{\Psi} \gamma^\mu D_\mu \Psi + \phi m \bar{\Psi}{\Psi} \right) \ ,
\end{align}
where $D_\mu = \partial_\mu + \omega^{(e)}_\mu$, and the spinor Lorentz connection here is the torsion-free metric-compatible connection constructed out of the tetrad. In the unimodular frame $\phi = e^{-\frac{1}{4}}$, we obtain
\begin{align}
S_{\mathrm{fermion}} = \int d^4x e \left( i\bar{\Psi} \gamma^\mu D_\mu \Psi + e^{-\frac{1}{4}} m \bar{\Psi}{\Psi} \right) \ ,
\end{align}
As we have emphasized, it is crucial to use the torsion-free metric-compatible connection here rather than the spin connection naturally defined in the (symmetric) teleparallel geometry. 

The other interaction among various fields such as gauge interactions and Yukawa interactions can be introduced with no difficulty.  
We have therefore shown that all the known matter couplings of the standard model of particle physics can be realized in our geometrical trinity of unimodular gravity.

\section{Discussions}
In this paper, we have formulated the teleparallel equivalent of unimodular gravity and the symmetric teleparallel equivalent of unimodular gravity in a unified framework. Our starting point is the Weyl invariant action with enhanced symmetry. The enhanced symmetry was important to guarantee that the resultant theory is equivalent to unimodular gravity or Einstein's general relativity with an adjustable cosmological constant.

Since all of them are physically equivalent, the difference may only reside in their interpretation. We may be able to find the superiority of one formulation over the other only if it is embedded in a deeper structure. For example, it is interesting to point out that the supergeometry in superspace is naturally equipped with torsion, so the Riemannian geometry that we are used to may not be the most natural formulation of gravity from the supersymmetric viewpoint.

Let us discuss a possible generalizations of $f(T,Q)$ type (symmetric) teleparallel gravity in unimodular gravity. The idea is to make the torsion tensor and the non-metricty tensor Weyl covariant:
\begin{align}
\hat{T}^a_{\ \mu\nu} &= \partial_\mu (\phi e^{a}_\nu) -  \partial_\nu (\phi e^{a}_\mu) \cr
&= \phi T^{a}_{\ \mu\nu} + (\partial_\mu \phi) e^a_\nu - (\partial_\nu \phi) e^a_\mu \cr 
\hat{Q}_{\alpha \mu\nu}&= \partial_\alpha(\phi^2 g_{\mu\nu}) \cr
& = \phi^2 Q_{\alpha \mu \nu} + 2 (\phi\partial_\alpha  \phi)  g_{\mu\nu} \ .
\end{align}
Then the theory constructed out of $\hat{T}^a_{\mu\nu}$ and $\hat{Q}_{\alpha \mu\nu}$ (as well as $\hat{g}_{\mu\nu} = \phi^2 g_{\mu\nu}$ and $\hat{e}^a_{\mu} = \phi e^{a}_\mu$) gives Weyl invariant action, which can be a starting point to construct the unimodular version of $f(T,Q)$ gravity (including new general relativity \cite{Hayashi:1979qx} and newer general relativity \cite{BeltranJimenez:2017tkd}) by setting $\phi= e^{-\frac{1}{4}}$.  Generic $f(T,Q)$ gravity lacks the enhanced symmetry and the consistency becomes non-trivial, and the unimodular version inherits the same difficulty, but there may exist consistent theories including extra propagating degrees of freedom.

In this paper, we have not discussed cosmological or astrophysical applications of these $f(T,Q)$ theories from our viewpoint. They are actively studied e.g. in \cite{Pfeifer:2022txm,Najera:2022jvm,Bahamonde:2022lvh,Paliathanasis:2022pgu,Albuquerque:2022eac,Duchaniya:2022rqu,DeBenedictis:2022sja,Bahamonde:2022ohm,Hohmann:2022wrk,Li:2022vtn,Koussour:2022rsv,Rigouzzo:2022yan,Huang:2022slc,Zhao:2022gxl,Kadam:2022lgq,Dimakis:2022rkd,Papanikolaou:2022hkg,Jusufi:2022loj,Gomes:2022vrc,Coley:2022qug,Briffa:2022fnv,Kadam:2022lxt,Tayde:2022lxd,Bahamonde:2022esv,Awad:2022fhx,Khyllep:2022spx,Paliathanasis:2022mrp,Lymperis:2022oyo,Bahamonde:2022zgj,Maurya:2022wwa,Calza:2022mwt,Mukherjee:2022yyq,Capozziello:2022agt}.\footnote{We have cited the papers that appeared in 2022. Earlier papers can be found in review articles such as \cite{Bahamonde:2021gfp}.}

\section*{Acknowledgements}
This work is in part supported by JSPS KAKENHI Grant Number 21K03581.


\begin{thebibliography}{99}

%\cite{Weinberg:1980kq}
\bibitem{Weinberg:1980kq}
S.~Weinberg and E.~Witten,
%``Limits on Massless Particles,''
Phys. Lett. B \textbf{96}, 59-62 (1980)
doi:10.1016/0370-2693(80)90212-9
%587 citations counted in INSPIRE as of 25 Aug 2022

%\cite{Alvarez:2006uu}
\bibitem{Alvarez:2006uu}
E.~Alvarez, D.~Blas, J.~Garriga and E.~Verdaguer,
%``Transverse Fierz-Pauli symmetry,''
Nucl. Phys. B \textbf{756}, 148-170 (2006)
doi:10.1016/j.nuclphysb.2006.08.003
[arXiv:hep-th/0606019 [hep-th]].
%131 citations counted in INSPIRE as of 19 Aug 2021

%\cite{Deser:1969wk}
\bibitem{Deser:1969wk}
S.~Deser,
%``Selfinteraction and gauge invariance,''
Gen. Rel. Grav. \textbf{1}, 9-18 (1970)
doi:10.1007/BF00759198
[arXiv:gr-qc/0411023 [gr-qc]].
%416 citations counted in INSPIRE as of 02 Sep 2022

%\cite{Alvarez:2005iy}
\bibitem{Alvarez:2005iy}
E.~Alvarez,
%``Can one tell Einstein's unimodular theory from Einstein's general relativity?,''
JHEP \textbf{03}, 002 (2005)
doi:10.1088/1126-6708/2005/03/002
[arXiv:hep-th/0501146 [hep-th]].
%96 citations counted in INSPIRE as of 21 May 2022

%\cite{Carballo-Rubio:2022ofy}
\bibitem{Carballo-Rubio:2022ofy}
R.~Carballo-Rubio, L.~J.~Garay and G.~Garc\'\i{}a-Moreno,
%``Unimodular Gravity vs General Relativity: A status report,''
[arXiv:2207.08499 [gr-qc]].
%0 citations counted in INSPIRE as of 25 Aug 2022






\bibitem{tele}
R.~Aldrovandi, J.G.~Pereira
Fundam.Theor.Phys. 173 (2013)

%\cite{Maluf:2013gaa}
\bibitem{Maluf:2013gaa}
J.~W.~Maluf,
%``The teleparallel equivalent of general relativity,''
Annalen Phys. \textbf{525}, 339-357 (2013)
doi:10.1002/andp.201200272
[arXiv:1303.3897 [gr-qc]].
%286 citations counted in INSPIRE as of 25 Aug 2022

%\cite{Bahamonde:2021gfp}
\bibitem{Bahamonde:2021gfp}
S.~Bahamonde, K.~F.~Dialektopoulos, C.~Escamilla-Rivera, G.~Farrugia, V.~Gakis, M.~Hendry, M.~Hohmann, J.~L.~Said, J.~Mifsud and E.~Di Valentino,
%``Teleparallel Gravity: From Theory to Cosmology,''
[arXiv:2106.13793 [gr-qc]].
%94 citations counted in INSPIRE as of 25 Aug 2022









%\cite{Nester:1998mp}
\bibitem{Nester:1998mp}
J.~M.~Nester and H.~J.~Yo,
%``Symmetric teleparallel general relativity,''
Chin. J. Phys. \textbf{37}, 113 (1999)
[arXiv:gr-qc/9809049 [gr-qc]].
%135 citations counted in INSPIRE as of 19 Aug 2021

%\cite{BeltranJimenez:2017tkd}
\bibitem{BeltranJimenez:2017tkd}
J.~Beltr\'an Jim\'enez, L.~Heisenberg and T.~Koivisto,
%``Coincident General Relativity,''
Phys. Rev. D \textbf{98}, no.4, 044048 (2018)
doi:10.1103/PhysRevD.98.044048
[arXiv:1710.03116 [gr-qc]].
%125 citations counted in INSPIRE as of 19 Aug 2021














%\cite{Jimenez:2019woj}
\bibitem{Jimenez:2019woj}
J.~B.~Jim\'enez, L.~Heisenberg and T.~S.~Koivisto,
%``The Geometrical Trinity of Gravity,''
Universe \textbf{5}, no.7, 173 (2019)
doi:10.3390/universe5070173
[arXiv:1903.06830 [hep-th]].
%110 citations counted in INSPIRE as of 19 Aug 2021


%\cite{BeltranJimenez:2019odq}
\bibitem{BeltranJimenez:2019odq}
J.~Beltr\'an Jim\'enez, L.~Heisenberg, D.~Iosifidis, A.~Jim\'enez-Cano and T.~S.~Koivisto,
%``General teleparallel quadratic gravity,''
Phys. Lett. B \textbf{805}, 135422 (2020)
doi:10.1016/j.physletb.2020.135422
[arXiv:1909.09045 [gr-qc]].
%39 citations counted in INSPIRE as of 19 Aug 2021



%\cite{Lu:2021wif}
\bibitem{Lu:2021wif}
J.~Lu, Y.~Guo and G.~Chee,
%``From GR to STG ---- Inheritance and development of Einstein's heritages,''
[arXiv:2108.06865 [gr-qc]].
%0 citations counted in INSPIRE as of 19 Aug 2021



%\cite{Nakayama:2021rda}
\bibitem{Nakayama:2021rda}
Y.~Nakayama,
%``Weyl transverse diffeomorphism invariant theory of symmetric teleparallel gravity,''
Class. Quant. Grav. \textbf{39}, no.14, 145006 (2022)
doi:10.1088/1361-6382/ac776b
[arXiv:2108.10465 [gr-qc]].
%3 citations counted in INSPIRE as of 25 Aug 2022






%\cite{Nakayama:2013is}
\bibitem{Nakayama:2013is}
Y.~Nakayama,
%``Scale invariance vs conformal invariance,''
Phys. Rept. \textbf{569}, 1-93 (2015)
doi:10.1016/j.physrep.2014.12.003
[arXiv:1302.0884 [hep-th]].
%245 citations counted in INSPIRE as of 19 Aug 2021



%\cite{Maluf:2011kf}
\bibitem{Maluf:2011kf}
J.~W.~Maluf and F.~F.~Faria,
%``Conformally invariant teleparallel theories of gravity,''
Phys. Rev. D \textbf{85}, 027502 (2012)
doi:10.1103/PhysRevD.85.027502
[arXiv:1110.3095 [gr-qc]].
%45 citations counted in INSPIRE as of 25 Aug 2022

%\cite{Bamba:2013jqa}
\bibitem{Bamba:2013jqa}
K.~Bamba, S.~D.~Odintsov and D.~S\'aez-G\'omez,
%``Conformal symmetry and accelerating cosmology in teleparallel gravity,''
Phys. Rev. D \textbf{88}, 084042 (2013)
doi:10.1103/PhysRevD.88.084042
[arXiv:1308.5789 [gr-qc]].
%132 citations counted in INSPIRE as of 25 Aug 2022


%\cite{Momeni:2014taa}
\bibitem{Momeni:2014taa}
D.~Momeni and R.~Myrzakulov,
%``Conformal Invariant Teleparallel Cosmology,''
Eur. Phys. J. Plus \textbf{129}, 137 (2014)
doi:10.1140/epjp/i2014-14137-8
[arXiv:1404.0778 [gr-qc]].
%15 citations counted in INSPIRE as of 25 Aug 2022


%\cite{Henneaux:1989zc}
\bibitem{Henneaux:1989zc}
M.~Henneaux and C.~Teitelboim,
%``The Cosmological Constant and General Covariance,''
Phys. Lett. B \textbf{222}, 195-199 (1989)
doi:10.1016/0370-2693(89)91251-3
%324 citations counted in INSPIRE as of 19 Aug 2021




%\cite{Edery:2019txq}
\bibitem{Edery:2019txq}
A.~Edery and Y.~Nakayama,
%``Palatini formulation of pure $R^2$ gravity yields Einstein gravity with no massless scalar,''
Phys. Rev. D \textbf{99}, no.12, 124018 (2019)
doi:10.1103/PhysRevD.99.124018
[arXiv:1902.07876 [hep-th]].
%31 citations counted in INSPIRE as of 19 Aug 2021


%\cite{BeltranJimenez:2022azb}
\bibitem{BeltranJimenez:2022azb}
J.~Beltr\'an Jim\'enez and T.~S.~Koivisto,
%``Lost in translation: The Abelian affine connection (in the coincident gauge),''
Int. J. Geom. Meth. Mod. Phys. \textbf{19}, no.07, 2250108 (2022)
doi:10.1142/S0219887822501080
[arXiv:2202.01701 [gr-qc]].
%7 citations counted in INSPIRE as of 28 Aug 2022



%\cite{Iosifidis:2018zwo}
\bibitem{Iosifidis:2018zwo}
D.~Iosifidis and T.~Koivisto,
%``Scale transformations in metric-affine geometry,''
Universe \textbf{5}, 82 (2019)
doi:10.3390/universe5030082
[arXiv:1810.12276 [gr-qc]].
%33 citations counted in INSPIRE as of 21 Aug 2021

%\cite{Gakis:2019rdd}
\bibitem{Gakis:2019rdd}
V.~Gakis, M.~Kr\v{s}\v{s}\'ak, J.~Levi Said and E.~N.~Saridakis,
%``Conformal gravity and transformations in the symmetric teleparallel framework,''
Phys. Rev. D \textbf{101}, no.6, 064024 (2020)
doi:10.1103/PhysRevD.101.064024
[arXiv:1908.05741 [gr-qc]].
%9 citations counted in INSPIRE as of 19 Aug 2021

%\cite{Hohmann:2021fpr}
\bibitem{Hohmann:2021fpr}
M.~Hohmann,
%``Variational Principles in Teleparallel Gravity Theories,''
Universe \textbf{7}, no.5, 114 (2021)
doi:10.3390/universe7050114
[arXiv:2104.00536 [gr-qc]].
%8 citations counted in INSPIRE as of 19 Aug 2021

%\cite{Alvarez:2010cg}
\bibitem{Alvarez:2010cg}
E.~Alvarez and R.~Vidal,
%``Weyl transverse gravity (WTDiff) and the cosmological constant,''
Phys. Rev. D \textbf{81}, 084057 (2010)
doi:10.1103/PhysRevD.81.084057
[arXiv:1001.4458 [hep-th]].
%19 citations counted in INSPIRE as of 19 Aug 2021


%\cite{Oda:2016pok}
\bibitem{Oda:2016pok}
I.~Oda,
%``Fake Conformal Symmetry in Unimodular Gravity,''
Phys. Rev. D \textbf{94}, no.4, 044032 (2016)
doi:10.1103/PhysRevD.94.044032
[arXiv:1606.01571 [gr-qc]].
%18 citations counted in INSPIRE as of 19 Aug 2021









%\cite{Hayashi:1979qx}
\bibitem{Hayashi:1979qx}
K.~Hayashi and T.~Shirafuji,
%``New General Relativity,''
Phys. Rev. D \textbf{19}, 3524-3553 (1979)
doi:10.1103/PhysRevD.19.3524
%920 citations counted in INSPIRE as of 26 Aug 2022





%\cite{Pfeifer:2022txm}
\bibitem{Pfeifer:2022txm}
C.~Pfeifer,
%``A quick guide to spacetime symmetry and symmetric solutions in teleparallel gravity,''
[arXiv:2201.04691 [gr-qc]].
%3 citations counted in INSPIRE as of 28 Aug 2022

%\cite{Najera:2022jvm}
\bibitem{Najera:2022jvm}
S.~N\'ajera, A.~Aguilar, G.~A.~Rave-Franco, C.~Escamilla-Rivera and R.~A.~Sussman,
%``Inhomogeneous solutions in $f(T,B)$ gravity,''
doi:10.1142/S0219887822400035
[arXiv:2201.06177 [gr-qc]].
%1 citations counted in INSPIRE as of 28 Aug 2022

%\cite{Bahamonde:2022lvh}
\bibitem{Bahamonde:2022lvh}
S.~Bahamonde, L.~Ducobu and C.~Pfeifer,
%``Scalarized black holes in teleparallel gravity,''
JCAP \textbf{04}, no.04, 018 (2022)
doi:10.1088/1475-7516/2022/04/018
[arXiv:2201.11445 [gr-qc]].
%5 citations counted in INSPIRE as of 28 Aug 2022

%\cite{Paliathanasis:2022pgu}
\bibitem{Paliathanasis:2022pgu}
A.~Paliathanasis and G.~Leon,
%``$f(T, B)$ gravity in a Friedmann-Lema\^\i{}tre-Robertson-Walker universe with nonzero spatial curvature,''
[arXiv:2201.12189 [gr-qc]].
%1 citations counted in INSPIRE as of 28 Aug 2022

%\cite{Albuquerque:2022eac}
\bibitem{Albuquerque:2022eac}
I.~S.~Albuquerque and N.~Frusciante,
%``A designer approach to f(Q) gravity and cosmological implications,''
Phys. Dark Univ. \textbf{35}, 100980 (2022)
doi:10.1016/j.dark.2022.100980
[arXiv:2202.04637 [astro-ph.CO]].
%12 citations counted in INSPIRE as of 28 Aug 2022

%\cite{Duchaniya:2022rqu}
\bibitem{Duchaniya:2022rqu}
L.~K.~Duchaniya, S.~V.~Lohakare, B.~Mishra and S.~K.~Tripathy,
%``Dynamical stability analysis of accelerating f(T) gravity models,''
Eur. Phys. J. C \textbf{82}, no.5, 448 (2022)
doi:10.1140/epjc/s10052-022-10406-w
[arXiv:2202.08150 [gr-qc]].
%3 citations counted in INSPIRE as of 28 Aug 2022

%\cite{DeBenedictis:2022sja}
\bibitem{DeBenedictis:2022sja}
A.~DeBenedictis, S.~Iliji\'c and M.~Sossich,
%``Spherically symmetric vacuum solutions and horizons in covariant f(T) gravity theory,''
Phys. Rev. D \textbf{105}, no.8, 084020 (2022)
doi:10.1103/PhysRevD.105.084020
[arXiv:2202.08958 [gr-qc]].
%1 citations counted in INSPIRE as of 28 Aug 2022

%\cite{Bahamonde:2022ohm}
\bibitem{Bahamonde:2022ohm}
S.~Bahamonde, K.~F.~Dialektopoulos, M.~Hohmann, J.~Levi Said, C.~Pfeifer and E.~N.~Saridakis,
%``Perturbations in Non-Flat Cosmology for $f(T)$ gravity,''
[arXiv:2203.00619 [gr-qc]].
%2 citations counted in INSPIRE as of 28 Aug 2022

%\cite{Hohmann:2022wrk}
\bibitem{Hohmann:2022wrk}
M.~Hohmann and C.~Pfeifer,
%``Gravitational wave birefringence in spatially curved teleparallel cosmology,''
[arXiv:2203.01856 [gr-qc]].
%2 citations counted in INSPIRE as of 28 Aug 2022

%\cite{Li:2022vtn}
\bibitem{Li:2022vtn}
M.~Li, Y.~Tong and D.~Zhao,
%``Possible consistent model of parity violations in the symmetric teleparallel gravity,''
Phys. Rev. D \textbf{105}, no.10, 104002 (2022)
doi:10.1103/PhysRevD.105.104002
[arXiv:2203.06912 [gr-qc]].
%1 citations counted in INSPIRE as of 28 Aug 2022

%\cite{Koussour:2022rsv}
\bibitem{Koussour:2022rsv}
M.~Koussour, S.~H.~Shekh and M.~Bennai,
%``Bianchi type-I Barrow holographic dark energy model in symmetric teleparallel gravity,''
[arXiv:2203.08181 [gr-qc]].
%1 citations counted in INSPIRE as of 28 Aug 2022

%\cite{Rigouzzo:2022yan}
\bibitem{Rigouzzo:2022yan}
C.~Rigouzzo and S.~Zell,
%``Coupling metric-affine gravity to a Higgs-like scalar field,''
Phys. Rev. D \textbf{106}, no.2, 2 (2022)
doi:10.1103/PhysRevD.106.024015
[arXiv:2204.03003 [hep-th]].
%1 citations counted in INSPIRE as of 28 Aug 2022

%\cite{Huang:2022slc}
\bibitem{Huang:2022slc}
Y.~Huang, J.~Zhang, X.~Ren, E.~N.~Saridakis and Y.~F.~Cai,
%``N-body simulations, halo mass functions, and halo density profile in $f(T)$ gravity,''
[arXiv:2204.06845 [astro-ph.CO]].
%1 citations counted in INSPIRE as of 28 Aug 2022

%\cite{Zhao:2022gxl}
\bibitem{Zhao:2022gxl}
Y.~Zhao, X.~Ren, A.~Ilyas, E.~N.~Saridakis and Y.~F.~Cai,
%``Quasinormal modes of black holes in $f(T)$ gravity,''
[arXiv:2204.11169 [gr-qc]].
%3 citations counted in INSPIRE as of 28 Aug 2022

%\cite{Kadam:2022lgq}
\bibitem{Kadam:2022lgq}
S.~A.~Kadam, B.~Mishra and J.~Said Levi,
%``Teleparallel scalar-tensor gravity through cosmological dynamical systems,''
Eur. Phys. J. C \textbf{82}, no.8, 680 (2022)
doi:10.1140/epjc/s10052-022-10648-8
[arXiv:2205.04231 [gr-qc]].
%0 citations counted in INSPIRE as of 28 Aug 2022


%\cite{Dimakis:2022rkd}
\bibitem{Dimakis:2022rkd}
N.~Dimakis, A.~Paliathanasis, M.~Roumeliotis and T.~Christodoulakis,
%``FLRW solutions in f(Q) theory: The effect of using different connections,''
Phys. Rev. D \textbf{106}, no.4, 043509 (2022)
doi:10.1103/PhysRevD.106.043509
[arXiv:2205.04680 [gr-qc]].
%3 citations counted in INSPIRE as of 28 Aug 2022


%\cite{Papanikolaou:2022hkg}
\bibitem{Papanikolaou:2022hkg}
T.~Papanikolaou, C.~Tzerefos, S.~Basilakos and E.~N.~Saridakis,
%``No constraints for $f(T)$ gravity from gravitational waves induced from primordial black hole fluctuations,''
[arXiv:2205.06094 [gr-qc]].
%4 citations counted in INSPIRE as of 28 Aug 2022

%\cite{Jusufi:2022loj}
\bibitem{Jusufi:2022loj}
K.~Jusufi, S.~Capozziello, S.~Bahamonde and M.~Jamil,
%``Testing Born-Infeld $f(T)$ teleparallel gravity through Sgr A$^\star$ observations,''
[arXiv:2205.07629 [gr-qc]].
%6 citations counted in INSPIRE as of 28 Aug 2022

%\cite{Gomes:2022vrc}
\bibitem{Gomes:2022vrc}
D.~A.~Gomes, J.~Beltr\'an Jim\'enez and T.~S.~Koivisto,
%``Energy and entropy in the Geometrical Trinity of gravity,''
[arXiv:2205.09716 [gr-qc]].
%3 citations counted in INSPIRE as of 28 Aug 2022

%\cite{Coley:2022qug}
\bibitem{Coley:2022qug}
A.~A.~Coley, R.~J.~v.~Hoogen and D.~D.~McNutt,
%``Symmetric Teleparallel Geometries,''
[arXiv:2205.10719 [gr-qc]].
%0 citations counted in INSPIRE as of 28 Aug 2022

%\cite{Briffa:2022fnv}
\bibitem{Briffa:2022fnv}
R.~Briffa, C.~Escamilla-Rivera, J.~Levi Said and J.~Mifsud,
%``$f(T,B)$ Gravity in the late Universe,''
[arXiv:2205.13560 [gr-qc]].
%1 citations counted in INSPIRE as of 28 Aug 2022

%\cite{Kadam:2022lxt}
\bibitem{Kadam:2022lxt}
S.~A.~Kadam, B.~Mishra and S.~K.~Tripathy,
%``Dynamical features of f(T,B) cosmology,''
Mod. Phys. Lett. A \textbf{37}, no.17, 2250104 (2022)
doi:10.1142/S0217732322501048
[arXiv:2206.00430 [gr-qc]].
%1 citations counted in INSPIRE as of 28 Aug 2022




%\cite{Tayde:2022lxd}
\bibitem{Tayde:2022lxd}
M.~Tayde, Z.~Hassan, P.~K.~Sahoo and S.~Gutti,
%``Static spherically symmetric wormholes in $f(Q,T)$ gravity,''
Chin. Phys. C \textbf{46}, 115101 (2022)
doi:10.1088/1674-1137/ac7f22
[arXiv:2206.01184 [gr-qc]].
%0 citations counted in INSPIRE as of 28 Aug 2022

%\cite{Bahamonde:2022esv}
\bibitem{Bahamonde:2022esv}
S.~Bahamonde, J.~Gigante Valcarcel, L.~J\"arv and J.~Lember,
%``Black hole solutions in scalar-tensor symmetric teleparallel gravity,''
[arXiv:2206.02725 [gr-qc]].
%1 citations counted in INSPIRE as of 28 Aug 2022

%\cite{Awad:2022fhx}
\bibitem{Awad:2022fhx}
A.~Awad, A.~Golovnev, M.~J.~Guzm\'an and W.~El Hanafy,
%``Revisiting diagonal tetrads: New Black Hole solutions in $f(T)$ gravity,''
[arXiv:2207.00059 [gr-qc]].
%0 citations counted in INSPIRE as of 28 Aug 2022


%\cite{Khyllep:2022spx}
\bibitem{Khyllep:2022spx}
W.~Khyllep, J.~Dutta, E.~N.~Saridakis and K.~Yesmakhanova,
%``Cosmology in $f(Q)$ gravity: A unified dynamical system analysis at background and perturbation levels,''
[arXiv:2207.02610 [gr-qc]].
%1 citations counted in INSPIRE as of 28 Aug 2022

%\cite{Paliathanasis:2022mrp}
\bibitem{Paliathanasis:2022mrp}
A.~Paliathanasis,
%``Anisotropic spacetimes in f(T,~B) theory I: Bianchi I universe,''
Eur. Phys. J. Plus \textbf{137}, no.8, 887 (2022)
doi:10.1140/epjp/s13360-022-03082-y
[arXiv:2207.08567 [gr-qc]].
%0 citations counted in INSPIRE as of 28 Aug 2022


%\cite{Lymperis:2022oyo}
\bibitem{Lymperis:2022oyo}
A.~Lymperis,
%``Late-time cosmology with phantom dark-energy in $f(Q)$ gravity,''
[arXiv:2207.10997 [gr-qc]].
%1 citations counted in INSPIRE as of 28 Aug 2022

%\cite{Bahamonde:2022zgj}
\bibitem{Bahamonde:2022zgj}
S.~Bahamonde and L.~J\"arv,
%``Coincident gauge for static spherical field configurations in symmetric teleparallel gravity,''
[arXiv:2208.01872 [gr-qc]].
%0 citations counted in INSPIRE as of 28 Aug 2022

%\cite{Maurya:2022wwa}
\bibitem{Maurya:2022wwa}
S.~K.~Maurya, K.~Newton Singh, S.~V.~Lohakare and B.~Mishra,
%``Anisotropic Strange Star Model Beyond Standard Maximum Mass Limit by Gravitational Decoupling in $f(Q)$ Gravity,''
[arXiv:2208.04735 [gr-qc]].
%1 citations counted in INSPIRE as of 28 Aug 2022


%\cite{Calza:2022mwt}
\bibitem{Calza:2022mwt}
M.~Calz\'a and L.~Sebastiani,
%``A class of static spherically symmetric solutions in $f(Q)$-gravity,''
[arXiv:2208.13033 [gr-qc]].
%0 citations counted in INSPIRE as of 02 Sep 2022

%\cite{Mukherjee:2022yyq}
\bibitem{Mukherjee:2022yyq}
P.~Mukherjee, J.~Levi Said and J.~Mifsud,
%``Neural Network Reconstruction of $H'(z)$ and its application in Teleparallel Gravity,''
[arXiv:2209.01113 [astro-ph.CO]].
%0 citations counted in INSPIRE as of 07 Sep 2022

%\cite{Capozziello:2022agt}
\bibitem{Capozziello:2022agt}
S.~Capozziello and M.~Shokri,
%``Slow-roll inflation in $f(Q)$ non-metric gravity,''
[arXiv:2209.06670 [gr-qc]].
%0 citations counted in INSPIRE as of 20 Sep 2022

\end{thebibliography}
\end{document}